      [1999/12/01 v1.4c Il Nuovo Cimento]
\documentclass{cimento}

\usepackage{graphicx}  
\title{Was the Universe neutral beyond redshift 6?}
\author{S.~Gallerani\from{ins:1}\from{ins:2}\ETC,
A.~Ferrara\from{ins:1},
X.~Fan\from{ins:3},
T.~Roy~Choudhury\from{ins:4},
R.~Salvaterra\from{ins:5}}
\instlist{\inst{ins:1} SISSA/International School for Advanced Studies, via Beirut 2-4, 34014 Trieste, Italy

\inst{ins:2} Institute of Physics, E\"{o}tv\"{o}s University, P\'{a}zm\'{a}ny P. s. 1/A, 1117 Budapest, Hungary 
\inst{ins:3} Steward Observatory, The University of Arizona, Tucson, AZ 85721, USA
\inst{ins:4} Institute of Astronomy, Madingley Road, Cambridge CB3 OHA, UK
\inst{ins:5} Dipartimento di Fisica G. Occhialini, Universita' degli studi di Milano Bicocca, Piazza della Scienza 3, I-20126 Milano, Italy

}
\begin{document}

\maketitle

\begin{abstract}
We provide measurements of the neutral hydrogen fraction $x_{\rm HI}$ at $z\sim 6$, by comparing semi-analytical models of the Ly$\alpha$ forest with observations of high-$z$ quasars and Gamma Ray Bursts absorption spectra.\\
We analyze the transmitted flux in a sample of 17 QSOs spectra at $5.74\leq z_{em}\leq 6.42$ studying separately the narrow transmission windows (peaks) and the wide dark portions (gaps) in the 
observed absorption spectra. By comparing the statistics of these spectral features with our models, we conclude that $x_{\rm{HI}}$ evolves smoothly from $10^{-4.4}$ at $z=5.3$ 
to $10^{-4.2}$ at $z=5.6$, with a robust upper limit $x_{\rm{HI}} < 0.36$ at $z=6.3$. We show the results of the first-ever detected transverse proximity effect in the HI Ly$\alpha$ forest, produced by the HII region of the faint quasar RD J1148+5253 
at $z=5.70$ intervening along the LOS of SDSS J1148+5251 at $z=6.42$.\\
Moreover, we propose a novel method to study cosmic reionization using absorption line spectra of high-redshift GRBs afterglows. We show that the 
time evolution and the statistics of gaps in the observed spectra represent exquisite tools to discriminate among different reionization models.  By applying our methods to GRB~050904 detected at $z=6.29$, we show that the observation of this burst
provides strong indications of a highly ionized intergalactic medium at $z\sim 6$, with an estimated mean neutral hydrogen fraction $x_{\rm HI}=6.4\pm 0.3 \times 10^{-5}$ along that line of sight.\\
\\
PACS 98.62.Ra - Intergalactic matter; quasar absorption systems; Lyman forest \\
PACS 98.54.Aj - Quasars\\
PACS 98.70.Rz - $\gamma$-ray bursts\\
\end{abstract}

\section{Introduction}
Although observations of cosmic 
epochs closer to the present have indisputably 
shown that the InterGalactic Medium (IGM) is in an ionized state, 
it is yet unclear when the phase transition from the neutral state to the 
ionized one started. Thus, the redshift of reionization, $z_{rei}$, is still very 
uncertain. 

 In the last few years, our knowledge of the reionization process has been enormously increased mainly 
owing to the observation of $z\sim 6$ QSOs by the SDSS survey \cite{fan06} and CMB
data \cite{page07}. Long gamma ray bursts (GRB) may constitute a complementary way 
to study the reionization process possibly probing even larger redshifts, the current recold holder being GRB~050904 at $z=6.3$ \cite{tagliaf05} \cite{haislip06}. 

We provide measurements of the neutral hydrogen fraction $x_{\rm HI}$ at epochs approaching reionization, by comparing semi-analytical models of the Ly$\alpha$ forest with observations of the highest-$z$ QSOs and GRBs absorption spectra.
\section{Ly$\alpha$ forest simulation}
The ultraviolet radiation emitted by a QSO/GRB can suffer resonant Ly$\alpha$ scattering as it propagates
through the intergalactic neutral hydrogen. In this process, photons are removed from the line of
sight (LOS)  resulting in an attenuation of the source flux, the so-called Gunn-Peterson (GP) 
effect.  To simulate the GP optical depth ($\tau_{GP}$) distribution 
we use the method described by Gallerani et al. (2006) \cite{galle06} and further revised in Gallerani et al. (2007) \cite{galle07}, whose main features are summarized as follows. 
Mildly non-linear density fluctuations giving raise to spectral absorption features in the 
intergalactic medium (IGM) can be described by a Log-Normal distribution. 
For a given IGM equation of state, the mean HI fraction, $x_{\rm HI}$, 
can be computed from photoionization equilibrium as a function of the baryonic overdensity, $\Delta\equiv\rho/\bar\rho$, and photoionization rate, $\Gamma$, due to the 
ultraviolet background radiation field. These quantities must be determined from a combination of theory and
observations; here we follow the approach of Choudhury \& Ferrara (2006) \cite{cf06}. This model contains two free parameters: 
(i) the star-formation efficiency $f_*$, and (ii) the escape fraction $f_{esc}$ of ionizing photons from galaxies. 
These are calibrated by a maximum-likelihood procedure to a broad observational data set. Currently, the available data can be explained by two different reionization histories, corresponding to different choices of the free parameters: 
(i) an Early Reionization Model (ERM) ($f_{*}=0.1; f_{esc}=0.07$), characterized by a highly 
ionized IGM  at $z>6$, and (ii) a Late Reionization Model (LRM) ($f_{*}=0.08; f_{esc}=0.04$), in which 
reionization occurs at $z\approx 6$. Both ERM and LRM by construction provide an excellent fit to the
mean neutral hydrogen fraction evolution experimentally deduced from the GP test.
\section{QSOs absorption spectra}
We first test the predictions of our models by applying various
statistical analysis to the simulated spectra and comparing our results with 
observations. Specifically, we use the following control statistics:
(i) Mean Transmitted Flux evolution in the redshift range $2-6$; 
(ii) Probability Distribution Function (PDF) of the transmitted flux 
at the mean redshifts $z=5.5, 5.7, 6.0$; (iii) Gap Width (GW) distribution in 
$3.5\leq z\leq 5.5$. For what concerns the GW statistics we define gaps as 
contiguous regions of the spectrum having a $\tau_{GP}> 2.5$ 
over rest-frame wavelength ($\lambda_{RF}$) intervals $> 1$~\AA. As both ERM and LRM successfully match 
the observational data at $z\leq 6$ for the control statistics considered, we 
proceed the comparison with more advanced statistical tools. 
\subsection{Largest Gap Width analysis}
\begin{figure}[h]
\centerline{
\includegraphics[width=8.8cm]{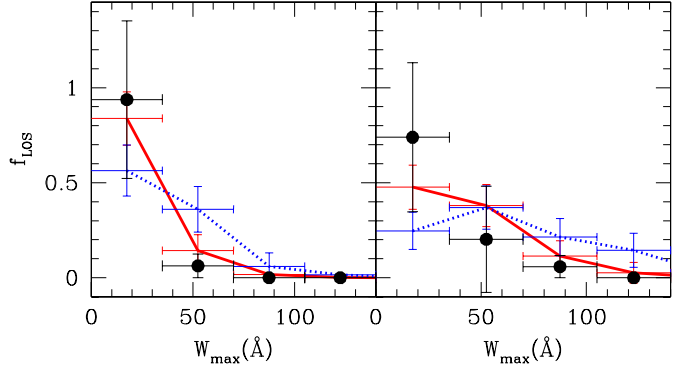}
}
\caption{Largest Gap Width distribution for the LR and the 
HR cases (left and right, respectively). 
Filled circles represent the result of the analysis of the 17 QSOs observed 
spectra.
Solid red (dotted blue) lines show the 
results obtained by the semi-analytical modeling implemented for the ERM 
(LRM). Vertical error bars measure poissonian noise, horizontal errors define 
the bin for the gap widths.}
\label{LGW}
\end{figure}
The Largest Gap Width (LGW) distribution quantifies the fraction of LOS 
which are characterized by the largest gap of a given width. We use observational data including 17 QSOs obtained by Fan et al. (2006).
We divide the observed spectra into two redshift-selected sub-samples: 
the ``Low-Redshift'' (LR) sample ($5.7 < z_{em} < 6$), and
the ``High-Redshift'' (HR) one ($ 6 < z_{em} < 6.4$). 
In order to measure  the evolution of 
$x_{\rm HI}$ with redshift, we apply the LGW both to simulated and observed spectra. From the comparison shown in Fig. 1 it results that both the LGW distributions predicted by the ERM and LRM provide a good fit to 
observational data. We exploit the agreement between the simulated and 
observed LGW distributions to derive an estimate of $x_{\rm HI}$. We find $\log_{10}x_{\rm HI}=-4.4^{+0.84}_{-0.90}$ at $z\approx 5.3$\footnote{The $x_{\rm HI}$ value 
quoted is the mean between the estimates predicted by the ERM and the LRM. 
Moreover, we consider the most conservative case in which the errors for the 
measurement of the neutral hydrogen fraction are provided by the minimum 
$x_{\rm HI}$ value found in the ERM and the maximum one in the LRM.}.
By applying the same method to the HR sample we constrain the 
neutral hydrogen fraction 
to be within $\log_{10}x_{\rm HI}=-4.2^{+0.84}_{-1.0}$ at $z\approx 5.6$.\\ 
Although the predicted LGW distributions are quite similar for the two models 
considered, yet some differences can be pointed out. Indeed in the HR case we 
find that a neutral hydrogen fraction at $z\approx 6$ 
higher than that one predicted by the LRM would imply an even worst 
agreement with observations, since a more abundant HI would produce a lower 
(higher) fraction of LOS characterized by the largest gap smaller (higher) 
than $40$ \AA\ with respect to observations. 
Thus, this study suggests $x_{\rm HI}<0.36$ at $z=6.32$ (obtained from the 
maximum value for $x_{\rm HI}$ found in the LRM at this epoch).

\subsection{Transverse proximity effect}
\begin{figure}[h]
\centerline{
\includegraphics[width=4.8cm]{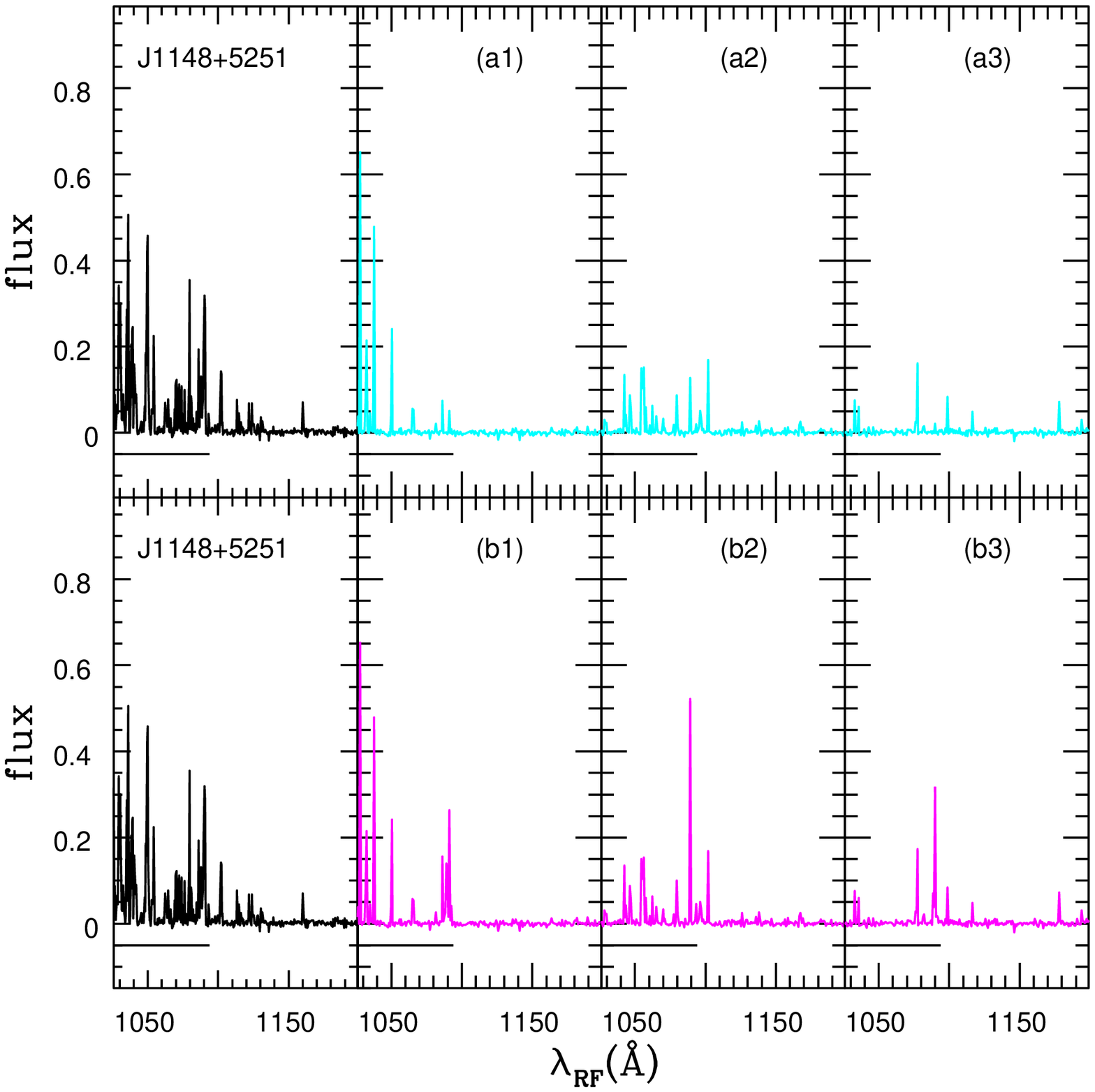}
\includegraphics[width=4.8cm]{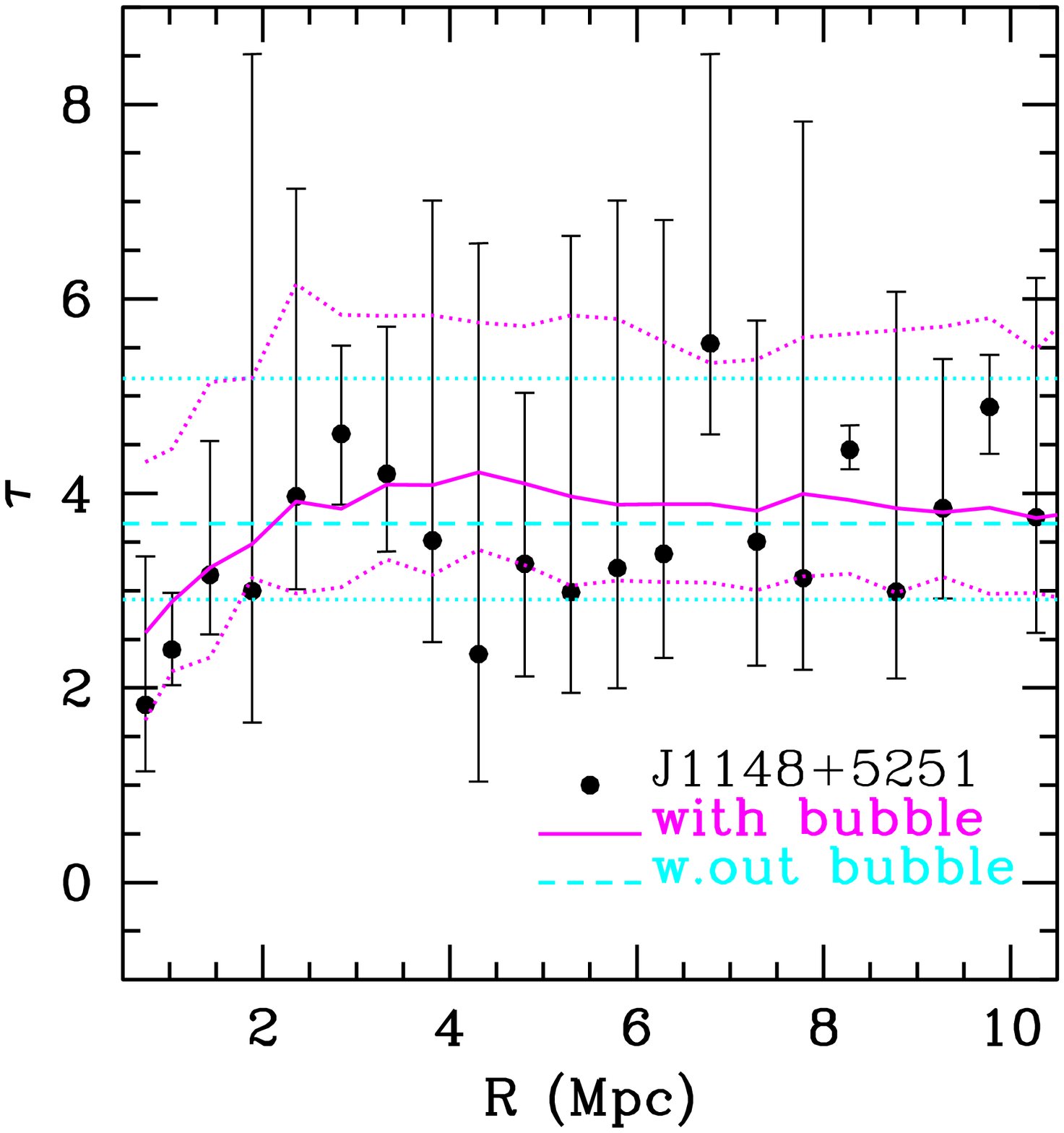}
}
\caption{{\bf Left} {\it Leftmost panels}: Observed transmitted flux (black spectra) 
in the spectrum of QSO SDSS J1148+5251 (QSO2, $z_{em}=6.42$). The solid black 
line shows the redshift path ($\Delta z_{prox}$) 
in which the bubble produced by 
QSO RD J1148+5252 (QSO1, $z_{em}=5.65$) intersects the LOS to QSO2.    
{\it Top panels (ai), with i=1,3}: 
Simulated fluxes (cyan spectra) along 3 different random LOS 
(cases ``without bubble''). {\it Bottom panels (bi), with i=1,3}: 
Simulated fluxes (magenta spectra) along the same LOSs 
shown in the top panels, taking into account the contribution from QSO1 to the 
total ionizing flux (cases ``with bubble''). {\bf Right} Evolution of the optical depth $\tau$ as a function of the distance R 
from QSO1. Filled circles denote the observed mean value for $\tau$, 
while error bars represent the maximum and the minimum observed $\tau$ at a 
given distance from the foreground QSO. Solid (dotted) magenta lines are the 
mean (maximum/minimum) values from 500 simulated LOS, computed adopting the 
case ``with bubble''. The dashed cyan horizontal line shows the mean optical 
depth predicted by the ERM in correspondence of the emission redshift of the 
foreground QSO. The dotted cyan horizontal lines denote the maximum/minimum 
optical depth at the same redshift.}
\end{figure}

Transmissivity windows in absorption spectra can be produced by ionizing 
sources whose bubbles intersect the lines of sight to the target object. Mahabal et al. (2005) \cite{maha05} have discovered a faint quasar 
(RD J1148+5253, hereafter QSO1) at $z=5.70$ in the field of the highest 
redshift quasar currently known (SDSS J1148+5251, hereafter QSO2) at $z=6.42$. 
In this Section we study the QSO2 transmitted flux, 
in order to analyze the proximity effect of QSO1 on the QSO2 spectrum.

In Fig. 2 (left panel) we compare the observed transmitted flux in the spectrum of QSO2 with the simulated fluxes along 
3 different LOS with (bottom row) or without (top) including the contribution from QSO1 to the total ionizing flux. 
For brevity, we refer to these case as ``with bubble'' or ``without bubble''. Visual inspection of Fig. 2
shows that the case ``with bubble'' is in better agreement with observations. Such statement can be made more 
quantitative by introducing a quantity denoted Peak Spectral Density (PSD), i.e. the number of peaks per
unit $\lambda_{RF}$ interval. For both the observed and simulated spectra, we compute the PSD 
inside and outside the bubble, finding the following results:\\
\centerline{
$(PSD_{obs}^{OUT},PSD_{obs}^{IN})=(0.11,0.46)$; $(PSD_{sim}^{OUT},PSD_{sim}^{IN})=(0.04^{+0.05}_{-0.04},0.22^{+0.32}_{-0.14}).$
}
Both in observations and simulations, the PSD is found to be  
$\approx 4.5$ times larger inside that bubble than outside it. 

As a final test for our model, we compute the observed evolution of the optical depth 
as a function of the distance $R$ from QSO1 and compare it with the predictions of model "with bubble'';
the result is shown in Fig. 2 (right panel). The agreement between observations 
and simulations is at 1-$\sigma$ confidence level for $70\%$ of the plotted 
points. For $R\leq 2$ Mpc, the mean optical depth $1.5 \leq \bar{\tau}\leq 3.5$ is 
lower than the mean value expected at $\bar{z}=5.65$ 
($\bar{\tau}_{5.65}\sim 4$); it approaches $\bar{\tau}_{5.65}$ at distances larger than $R_{\tau}\sim 2$ Mpc. By taking the difference between $R_{\tau}$ and $R_{\bot}$, we set a lower limit on the foreground QSO lifetime $t_Q>\frac{R_{\tau}-R_{\bot}}{c}+(t_{\tau}-t_{QSO1})\approx 11$ Myr, where $t_{\tau}$ and $t_{QSO1}$ represent the cosmic times corresponding to the redshifts $z_{\tau}=5.68$ and $z_{em}^{QSO1}=5.65$, respectively.

The physical interpretation of the results reported in this Section is the following. In the case ``with bubble'', in correspondence of spectral regions where
$\Gamma_{\rm HI}^{QSO1}\geq \Gamma_{\rm HI}$, most of the gaps present in 
the case ``without bubble'' disappear, making room for peaks, 
as a consequence of the decreased opacity in the proximity of QSO1.  
The enhancement in the transmissivity decreases moving toward outside the bubble, as $\Gamma_{\rm HI}^{QSO1}<\Gamma_{\rm HI}$. 
This is the first-ever detection of the transverse proximity effect in the HI Ly$\alpha$ forest. 
\section{GRBs absorption spectra}
We have built a database of synthetic GRB afterglow emission spectra starting from
the observed spectral energy distribution and time evolution of the most
distant GRB detected up-to-now, i.e. GRB 050904 \cite{tagliaf05} \cite{kawai06}. The unabsorbed afterglow spectrum of GRB 050904 can be
parameterized as $F(\nu)\propto \nu^{\rm \alpha} t^{\rm \beta}$, where we assume for $\alpha$ and $\beta$ the values founded by previous studies \cite{tagliaf05} \cite{haislip06}. Finally, we normalize the intrinsic GRB 050904 
optical spectrum in order to reproduce the flux of 
$\sim 18$ $\mu$Jy as measured at 1 day from burst in the J band \cite{haislip06}. 
We simulate the observed flux $F_{\rm obs}=F(\nu)~e^{-\tau_{\rm GP}}$ of GRBs absorption spectra in the rest frame spectrum between Ly$\alpha$ ($1215.67$ \AA) and Ly$\beta$ ($1025.72$ \AA) at different times after the burst.

\begin{figure}[h]
\centerline{
\includegraphics[width=8.8cm]{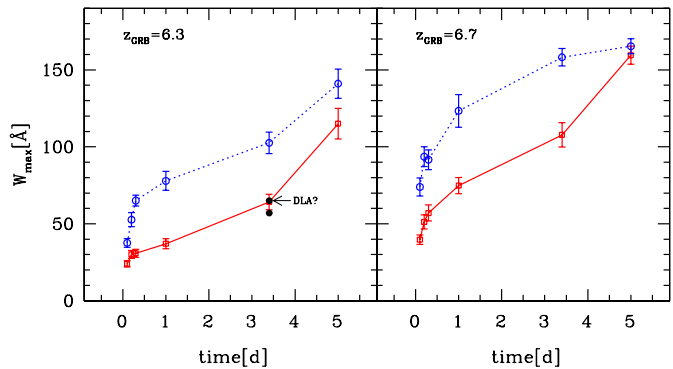}     
}
\caption{Evolution of the largest transmissivity gap found in the rest frame 
GRB afterglow spectra as a function of the observer time after the explosion for the two selected 
GRB redshifts $z=6.3$ (left panel) and $z=6.7$ (right). The red solid (blue dashed) line refers
to Early (Late) Reionization Model, with the error bars showing the standard deviation associated to the weighted mean computed from 10 realizations of 5 LOS. The black filled circle refers to the largest gap
measured in the spectrum of GRB~050904 afterglow.}
\end{figure}
\subsection{Largest Gap Width analysis}
The main idea we propose is to exploit the statistics of the transmissivity gaps imprinted by the
intervening IGM neutral hydrogen on the otherwise smooth power-law spectrum of high-redshift GRBs. On general grounds, we 
expect that at any given redshift, but particularly above $z=6$, where differences become more marked, the value of 
$x_{\rm{HI}}$ is higher in the LRM than in the ERM. As a result,  wider and more numerous gaps are expected 
if reionization completes later. Moreover, as the time after the burst increases, the gaps become larger.
In fact, the progressive fading of the unabsorbed afterglow produces a
corresponding attenuation of the observed flux.\\
To put the above arguments on more quantitative grounds, we have derived, using the procedure described above, the evolution 
of the LGW  found in synthetic afterglow spectra with time after explosion. The results of the calculation are shown in Fig. 3 for two selected GRB redshifts $z_{\rm GRB}=6.3, 6.7$ and for the ERM and LRM cases.     
The differences caused by the two different reionization histories are striking. Since the beginning, the width of the gaps is a factor $\approx 2$  times wider in the LRM
(38~\AA~vs 24~\AA~at $z_{\rm GRB}=6.3$, 74~\AA~vs 40~\AA~at $z_{\rm GRB}=6.7$) than in the ERM. 
We compare these predictions to the results of the analysis of GRB 050904 spectrum obtained 3.4 days after the burst (Kawai et al. 2006), whose largest dark gap is 
$W_{\rm max}\sim 65$~\AA~in the source rest frame. This value refers to the dark region immediately blueward the Ly$\alpha$ emission line ($\sim [8420; 8880]$).  
The black filled circle in Fig. 3, 
corresponding to the values of $W_{\rm max}$ and $t$ for GRB 050904, clearly shows that the ERM is favored by the data. In the LRM the typical $W_{\rm max}$ at $t=3.4$~days is as large as $\sim$ 100~\AA, well 
above the observed value. \\
We derive the mean neutral hydrogen fraction $x_{\rm HI}$ along the synthetic lines of sight characterized by a LGW in the interval $65\pm 5$~\AA, centered on the size of the largest dark gap observed in the GRB~050904 spectrum. The observed LGW in the GRB~050904 afterglow spectrum is consistent with $x_{\rm HI}=6.4\pm 0.3\times 10^{-5}$ \cite{galle07b}. This result is in agreement with previous measurements by Totani et al. (2006), who find that $x_{\rm HI}$ is consistent with zero with upper limit $x_{\rm HI}<0.17$ at 68\% C.L.
\section{Conclusions}
We measure the neutral hydrogen fraction $x_{\rm HI}$ at epochs approaching the reionization, by comparing semi-analytical models of the Ly$\alpha$ forest with observations of high-$z$ quasars and Gamma Ray Bursts absorption spectra. We consider an Early Reionization Model (ERM), characterized by a highly ionized Universe at $z\sim 6$ and a Late Reionization Model (LRM) in which 
reionization occurs at $z\sim 6$.\\
By comparing statistical analysis of the transmitted flux in a sample of 17 QSOs spectra at $5.74\leq z_{em}\leq 6.42$ with our models, we find that both ERM and LRM provide good fits to the 
observed LGW distribution, favoring a scenario
in which $x_{\rm HI}$ smoothly evolves from $10^{-4.4}$ at $z\approx 5.3$ to 
$10^{-4.2}$ at $z\approx 5.6$, with a robust upper limit $x_{\rm{HI}} < 0.36$ at $z=6.3$. Discriminating among the two reionization scenarios would require a sample of QSO at even higher redshifts.\\
We show the results of the first-ever detected transverse proximity effect in the HI Ly$\alpha$ forest, produced by the HII region of the faint quasar RD J1148+5253 
at $z=5.70$ intervening along the LOS of SDSS J1148+5251 at $z=6.42$.\\
Moreover, we show that the time evolution of gaps in GRBs absorption spectra represent exquisite tools to discriminate among different reionization models.  By applying our methods to GRB~050904 detected at $z=6.29$, we show that the observation of this burst
provides strong indications of a highly ionized intergalactic medium at $z\sim 6$, with an estimated mean neutral hydrogen fraction $x_{\rm HI}=6.4\pm 0.3 \times 10^{-5}$ along that line of sight.

\end{document}